\documentclass[floatfix,aps,prd,showpacs,amsmath,nofootinbib,preprintnumbers,twocolumn]{revtex4}

\usepackage{graphicx}
\usepackage{bm}

\newcommand{\figurewidth}{3.in}

\def\({\left(}
\def\){\right)}
\def\[{\left[}
\def\]{\right]}

\def\e{\begin{equation}}
\def\q{\end{equation}}
\def\m{\begin{eqnarray}}
\def\n{\end{eqnarray}}

\begin{document}

\title{The Dark Side of the Universe after Planck}

\author{Cheng Cheng$^{1,2}$\footnote{chcheng@itp.ac.cn} and Qing-Guo Huang$^1$\footnote{huangqg@itp.ac.cn}}
\affiliation{
$^1$ State Key Laboratory of Theoretical Physics, Institute of Theoretical Physics, 
Chinese Academy of Science, Beijing 100190, China \\
$^2$ University of the Chinese Academy of Sciences, Beijing 100190, China}

\date{\today}

\begin{abstract}

Recently released Planck data implies a smaller Hubble constant $H_0$ than that from Hubble Space Telescope project (HST) and a larger percentage of the matter components $\Omega_m$ compared to Supernova Legacy Survey (SNLS) in $\Lambda$CDM model. In this paper we found that even though the tension on $H_0$ between Planck and HST can be relaxed if the dark radiation is introduced ($\Delta N_{\rm eff}=0.536_{-0.224}^{+0.229}$ at $68\%$ CL from the datasets of Planck+WMAP Polarization (WP)+baryon acoustic oscillation (BAO)+the combination of supernova Union2.1 compilation of 580 SNe (Union2.1)+HST), $\Omega_m$ from Planck is still not nicely compatible with that from SNLS. The tensions between Planck and other astrophysical datasets can be significantly relaxed in $w$CDM model, and the combination of these datasets prefers a phantom-like dark energy at more than $95\%$ CL: $w=-1.15\pm 0.07$ and $w=-1.16\pm 0.06$ at $68\%$ CL from Planck+WP+BAO+Union2.1+HST and Planck+WP+BAO+SNLS+HST respectively. From the statistical point of view, there is no evidence for a time-evolving equation of state ($\Delta \chi^2=-0.3$ compared to a constant equation of state for the combination of Planck+WP+BAO+SNLS+HST).

\end{abstract}


\maketitle


\section{Introduction}

By now, the existence of dark matter is well accepted.
In addition, cosmic acceleration discovered in 1998 \cite{Riess:1998cb,Perlmutter:1998np} is one of the most important discoveries in the last century. The present cosmic acceleration can be driven by the dark energy due to its negative pressure.
Besides the dark matter and dark energy, some unknown relativistic species, so-called dark radiation, may also exist \cite{Hinshaw:2012aka,Hou:2012xq}. 

The total energy density of radiation in the Universe is 
\m
\rho_{\rm rad}=\[1+{7\over 8}\({4\over 11}\)^{4\over 3}N_{\rm eff} \]\rho_\gamma, 
\n
which is a sum of the CMB photon energy density $\rho_\gamma={\pi^2\over 15}T^4$, the energy density in standard model (SM) neutrinos with $N_{\rm eff}^{\rm SM}=3.046$ and any departure from the standard scenario parametrized as a summand in $N_{\rm eff}=3.046+\Delta N_{\rm eff}$. A positive  value of $\Delta N_{\rm eff}$ implies that there is dark radiation in the Universe. If there is any dark radiation, it changes the energy budget at recombination and has significant imprint in the anisotropies of cosmic microwave background (CMB).  But the dark radiation becomes ignorable in the late time Universe, and then the dark radiation does not affect the fitting to the data of baryon acoustic oscillation (BAO) \cite{bao}, Supernova Legacy Survey (SNLS) samples \cite{Conley:2011ku}, the combination of supernova Union2.1 compilation of 580 SNe (Union2.1) \cite{Suzuki:2011hu} and Hubble Space Telescope (HST) project \cite{Riess:2011yx}.

Recently the first cosmological results \cite{Ade:2013zuv} from the Planck were released. Planck is a very accurate experiment to measure the tiny anisotropies in the cosmic microwave background radiation (CMBR). From the data of Planck ones can learn about the composition and evolution of the Universe from its birth to the present day. Even though the standard six parameter $\Lambda$CDM model is preferred by Planck data, some tensions between Planck and some other astrophysical datasets are also intriguing. Here we mainly focus on two significant tensions: \\
$\bullet$ Even though the BAO data excellently agrees with Planck in the base $\Lambda$CDM model, the $H_0$ prior from HST is 
\m
H_0=(73.8\pm 2.4)\ {\rm km}\ {\rm s}^{-1}\ {\rm Mpc}^{-1}\ (68\%\ {\rm CL};\ {\rm HST})
\n
which is discrepant with the Planck estimate in the base $\Lambda$CDM model at about $2.5\ \sigma$ level. Note that the constraint on $H_0$ from Planck+WMAP Polarization (WP) is 
\m
H_0=(67.3\pm 1.2)\ {\rm km}\ {\rm s}^{-1}\ {\rm Mpc}^{-1}\ (68\%\ {\rm CL};\ {\rm Planck+WP}). 
\n 
See Section 5.3 of \cite{Ade:2013zuv} in detail. \\
$\bullet$ Even though there is no obvious inconsistency between the SNe samples, the posterior distribution for $\Omega_m$ from Planck+WP in the base $\Lambda$CDM model is in some tension with the distribution from SNLS. See Section 5.4 of \cite{Ade:2013zuv} in detail.

Here we will investigate how the dark components in the Universe can help us to tackle the tensions between Planck and some other astrophysical datasets. Methods and datasets used in this paper is described in Secion II. In Section III, we separately consider three dark energy models, i.e. $\Lambda$CDM, $w$CDM and (CPL)CDM model, with/without dark radiation. And discussion is included in Section IV. 
Our main results are summarised in Table.~\ref{tab:sum}. 
\begin{table*}[htbp]
\centering
\renewcommand{\arraystretch}{1.5}
\scriptsize 
{
 
\

\begin{tabular}{c|cc|cc}
\hline\hline
Parameters & \multicolumn{2}{|c|}{Planck+WP+BAO+Union2.1+HST} & \multicolumn{2}{c}{Planck+WP+BAO+SNLS+HST} \\
\hline
$\Omega_bh^2$&$0.02246_{-0.00025}^{+0.00026}$  &$0.02202\pm{0.00026}$  &$0.02202_{-0.00026}^{+0.00025}$&$0.02193_{-0.00026}^{+0.00027}$\\
$\Omega_ch^2$&$0.1266\pm{0.0042}$  &$0.1207_{-0.0022}^{+0.0021}$  &$0.1207\pm{0.0021}$ & $0.1220\pm{0.0025}$ \\
$100\theta_{\rm MC}$&$1.04071_{-0.00067}^{+0.00068}$  &$1.04122\pm{0.00060}$ &$1.04122_{-0.00059}^{+0.00058}$&$1.04102\pm{0.00061}$\\
$\tau$&$0.0958_{-0.0133}^{+0.0134}$ &$0.0879_{-0.0126}^{+0.0127}$ &$0.0882\pm{0.0125}$&$0.0861_{-0.0127}^{+0.0126}$ \\
${\rm{ln}}(10^{10}A_s)$&$3.12\pm{0.028}$ &$3.22_{-0.028}^{+0.029}$ &$3.09_{-0.025}^{+0.024}$ &$3.09\pm{0.024}$\\
$n_s$&$0.9805\pm{0.0080}$&$0.9587_{-0.0063}^{+0.0064}$ &$0.9584\pm{0.0063}$&$0.9556_{-0.0069}^{+0.0070}$\\
\hline
$w_0$&$-1$&$-1.15\pm 0.07$  &$-1.16\pm 0.06$&$-1.05\pm 0.15$\\
$w_a$& -  &- &-&$-0.596_{-0.723}^{+0.717}$\\
$\Delta N_{\rm eff}$&$0.536_{-0.224}^{+0.229}$&-&- &- \\
\hline
$H_0$&$71.23_{-1.37}^{+1.36}$ &$71.28\pm{1.49}$ &$71.65\pm{1.33}$ &$71.69\pm{1.37}$\\
$\Omega_m$&$0.295\pm{0.009}$&$0.283\pm 0.012$&$0.279_{-0.010}^{+0.011}$&$0.282\pm{0.011}$ \\
\hline
$-\ln$ (like) & $4968.69$ &$4968.79$  &$5114.43$ & $5114.28$ \\
\hline
\end{tabular}
}
\caption{Cosmological Parameters. The dark energy equation of state is parametrized by $w(a)=w_0+w_a(1-a)$. The cosmological constant corresponds to $w_0=-1$ and $w_a=0$, $w$CDM model corresponds to $w_a=0$.  }
\label{tab:sum}
\end{table*}

\section{Methodology}
The 6-parameter $\Lambda$CDM model is taken as the base model in this paper and the pivot scale is set as $k_0 = 0.002 \hbox{Mpc}^{-1}$. The six free-running parameters are $\{$$\Omega_bh^2$, $\Omega_ch^2$, $\theta$, $\tau$, $n_s$, $A_s$$\}$, which represent physical baryon density, physical cold dark matter density, the ratio of the angular diameter distance to the LSS sound horizon, optical depth, spectral index and amplitude of primordial scalar perturbation. $N_{\rm eff}$ describing the effective number of massless neutrinos is introduced in the extended model which can include dark radiation. We reformulate it as $N_{\rm eff}=3.046+\Delta N_{\rm eff}$ to investigate its departure from the standard scenario. A positive value of $\Delta N_{\rm eff}$ indicates the existence of dark radiation. 


We fit these parameters with the datasets of Planck, WP, BAO, SNLS, Union2.1 and HST by calculating the power spectrum with CAMB \cite{CAMB} and running the Monte Carlo Markov Chain(MCMC) with CosmoMC pacage\cite{cosmomc}. The likelihood used here is provided by Planck group \cite{Ade:2013zuv,plancklikelihood} and CosmoMC pacage \cite{cosmomc}.

\section{Results}
\subsection{$\Lambda$CDM (+$\Delta N_{\rm eff}$) model}

\begin{figure}[ht]
\centerline{\includegraphics[width=\figurewidth]{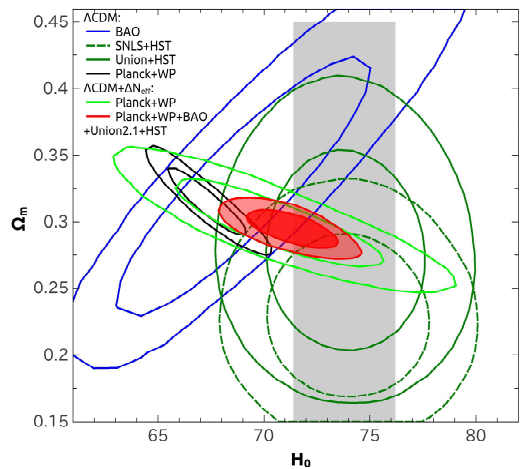}}
\caption{(color online). Contour plots of $\Omega_m - H_0$ in $\Lambda$CDM (+$\Delta N_{\rm eff}$) model with different datasets in $68\%$ and $95\%$ confidence region. The light grey band indicates the prior of $H_0$ from HST project at $1 \sigma$ level. 
The black, blue, olive solid and olive dashed contours enclose the $68\%$ and $95\%$ confidence regions from Planck+WP, BAO, Union2.1+HST and SNLS+HST in the base $\Lambda$CDM model. The red shaded region and green  contour respectively enclose the $68\%$ and $95\%$ confidence regions from Planck+WP, and Planck+WP+BAO+Union2.1+HST in the $\Lambda$CDM+$\Delta N_{\rm eff}$ model respectively.
}
\label{fig:lcdm}
\end{figure}

A cosmological constant $\Lambda$ is the simplest candidate for dark energy. The equation of state for cosmological constant is $w\equiv {p_{\rm DE}/\rho_{\rm DE}}=-1$. However, there are some tensions between Planck and other astrophysical data, such as HST and SNLS, in the base $\Lambda$CDM model. See the discrepancy between the black and olive dashed contours in Fig.~\ref{fig:lcdm} where the black and olive dashed contours indicate the constraints from Planck+WP and SNLS+HST  respectively. 

\begin{figure}[ht]
\centerline{\includegraphics[width=\figurewidth]{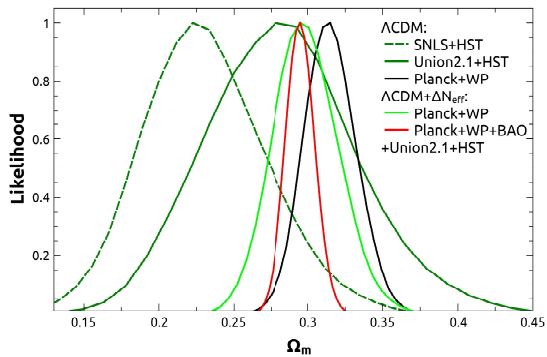}}
\caption{(color online). Distribution of $\Omega_m$ from different datasets and models. 
The black, olive solid and olive dashed curves indicate the distributions of $\Omega_m$ from Planck+WP, Union2.1 and SNLS in the base $\Lambda$CDM model. The green and red curves are the distributions from Planck+WP, and Planck+WP+BAO+Union2.1+HST in the $\Lambda$CDM+$\Delta N_{\rm eff}$ model.}
\label{fig:lcdmm}
\end{figure}

The marginalised constraint on $\Omega_m$ from the combination of SNLS and HST is 
\m
\Omega_m=0.231_{-0.038}^{+0.039}\ (68\%\ {\rm CL};\ {\rm SNLS+HST}). 
\n
On the other hand, the constraint on $\Omega_m$ from Union2.1 and HST is given by 
\m
\Omega_m=0.280\pm 0.048 \ (68\%\ {\rm CL};\ {\rm Union2.1+HST}). 
\n
One can see that there is no tension between different SNIa datasets, such as SNLS and Union2.1 (see the olive solid and dashed curves in Fig.~\ref{fig:lcdmm}). 

\begin{figure*}[ht]
\centerline{\includegraphics[width=\figurewidth]{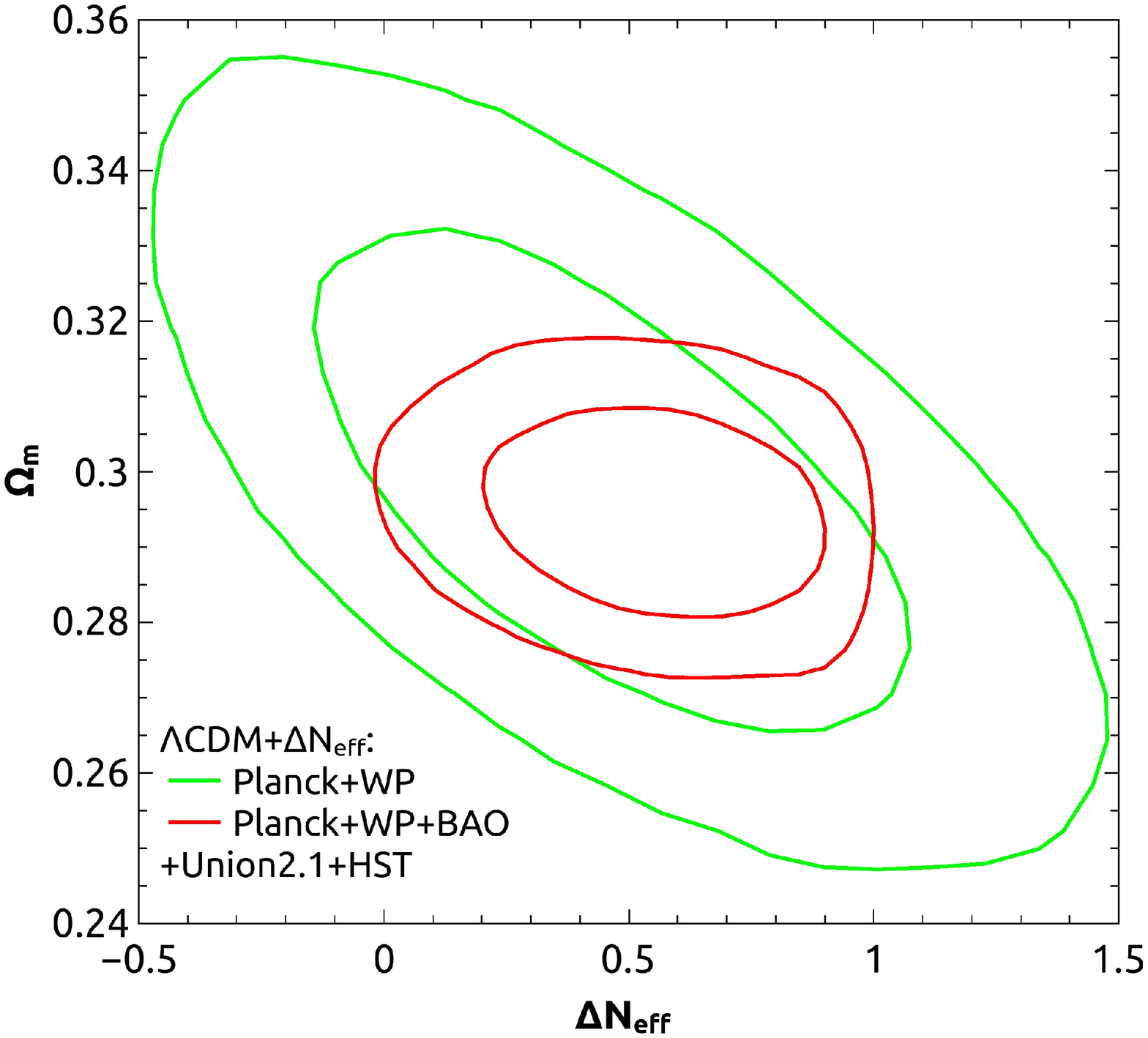}\quad \includegraphics[width=\figurewidth]{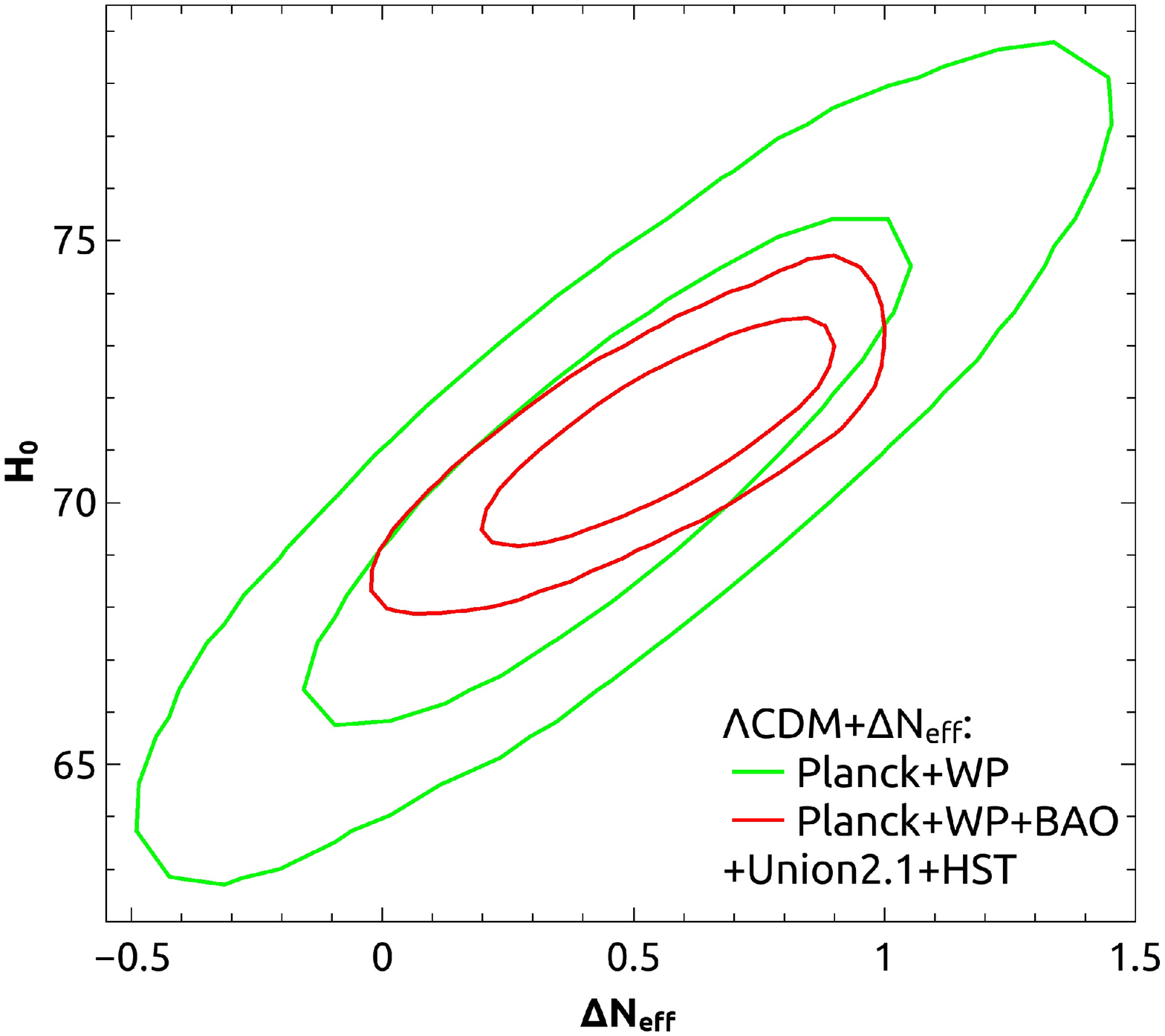}}
\caption{(color online). Contour plots of $\Delta N_{\rm eff} - \Omega_m$ and $\Delta N_{\rm eff} - H_0$ from different datasets in the $\Lambda$CDM+$\Delta N_{\rm eff}$ model. The green and red contours correspond to Planck+WP and Planck+WP+BAO+Union2.1+HST respectively.}
\label{fig:lcdmneff}
\end{figure*}

In this subsection we consider to add a new component, so-called dark radiation, in the energy budget. Since the dark radiation can change the energy budget at recombination, one can expect that it will change the constraints on $\Omega_m$ and $H_0$ from CMB data (Planck+WP). See the correlations related to $\Delta N_{\rm eff}$ in Fig.~\ref{fig:lcdmneff}. 
In $\Lambda$CDM+$\Delta N_{\rm eff}$ model, the combination of Planck and WP implies 
\m
\Omega_m=0.299\pm 0.021, 
\n
and 
\m
H_0=(70.7_{-3.1}^{+3.2})\ {\rm km}\ {\rm s}^{-1}\ {\rm Mpc}^{-1}, 
\n
at $68\%$ CL. See the green contour in Fig.~\ref{fig:lcdm}. Compared to the black contour in Fig.~\ref{fig:lcdm}, the green one covers a larger region and it indicates that the tension between CMB data and the $H_0$ prior from HST project can be significantly relaxed. This was also pointed out by Planck collaboration in \cite{Ade:2013zuv}. 
On the other hand, comparing olive dashed curve with the green curve in Fig.~\ref{fig:lcdmm}, we find that SNLS still prefers a smaller value of $\Omega_m$ even in the $\Lambda$CDM+$\Delta N_{\rm eff}$ model. The measurement of SNLS is not compatible with Planck+WP estimate at about $1.7\sigma$.


From Fig.~\ref{fig:lcdm}, we find that there is an overlap among Planck+WP, Union2.1+HST and BAO in the $\Lambda$CDM+$\Delta N_{\rm eff}$ model. 
Therefore we can combine all of them to constrain $\Omega_m$, $H_0$ and $\Delta N_{\rm eff}$, namely 
\m
H_0=(71.2\pm 1.4)\ {\rm km}\ {\rm s}^{-1}\ {\rm Mpc}^{-1}, 
\n
\m
\Omega_m=0.295\pm 0.009,  
\n
and 
\m
\Delta N_{\rm eff}=0.536_{-0.224}^{+0.229}, 
\n
at $68\%$ CL. See the red shaded contour plot in Fig.~\ref{fig:lcdm}. It implies that there should be dark radiation at around $2.4 \sigma$ level. 
Some other  investigations about the dark radiation has also been done after Planck 2013 data release.  In \cite{Said:2013hta} a presence of a dark radiation component is preferred at $91.1 \%$ C.L. from Planck+WP if the power spectrum amplitude $A_L$ is also allowed to vary. Combining with the matter power spectrum from BOSS DR9 and the prior on the Hubble constant from HST, Planck dataset implies a non standard number of neutrino species at $2.3 \sigma$ and the upper bound on the sum of neutrino masses is $0.51$ eV at $95\%$ C.L. in \cite{Archidiacono:2013fha}. The authors adopted Bayesian model comparison to determine the number of relativistic species fro Planck data, but did not find any evidence for deviations from the standard cosmological model in \cite{Verde:2013cqa}. And the still allowed parameter space in sterile neutrino models, hadronic axion models were explored in \cite{Valentino:2013wha}. 

\begin{figure}[ht]
\centerline{\includegraphics[width=\figurewidth]{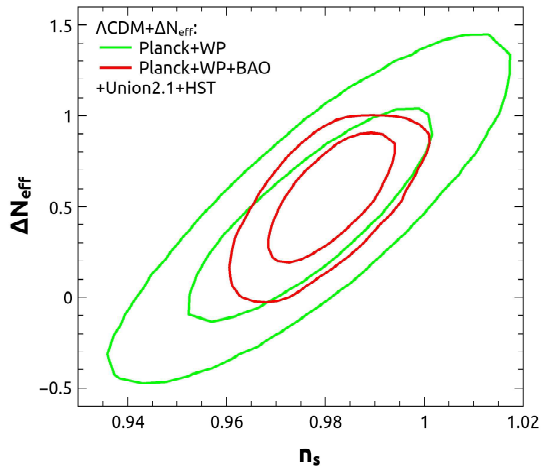}}
\caption{(color online). Plot of $\Delta N_{\rm eff} - n_s$ from different datasets in the $\Lambda$CDM+$\Delta N_{\rm eff}$ model. The green and red contours correspond to Planck+WP and Planck+WP+BAO+Union2.1+HST respectively.}
\label{fig:lcdmnsneff}
\end{figure}

In addition, we find that there is correlation between $\Delta N_{\rm eff}$ and the spectral index $n_s$, because the dark radiation can suppress the high-$\ell$ power spectrum and a redder-tilted primordial power can make a similar effect on the high-$\ell$ power spectrum. Such correlation is showed in Fig.~\ref{fig:lcdmnsneff}.
Larger $\Delta N_{\rm eff}$, larger $n_s$. In the combination of Planck+WP+BAO+Union2.1+HST, the constraint on $n_s$ becomes $n_s=0.9805\pm 0.0080$ which implies that there is no highly significant deviation from scale-invariance of the primordial power spectrum (only at around $2.4\sigma$ level). 

We conclude that even though the tension on $H_0$ between HST and Planck+WP can be significantly relaxed once the dark radiation is taken into account, $\Omega_m$ from Planck is still not nicely compatible with that from SNLS.


\subsection{$w$CDM (+$\Delta N_{\rm eff}$) model}

\begin{figure}[ht]
\centerline{\includegraphics[width=\figurewidth]{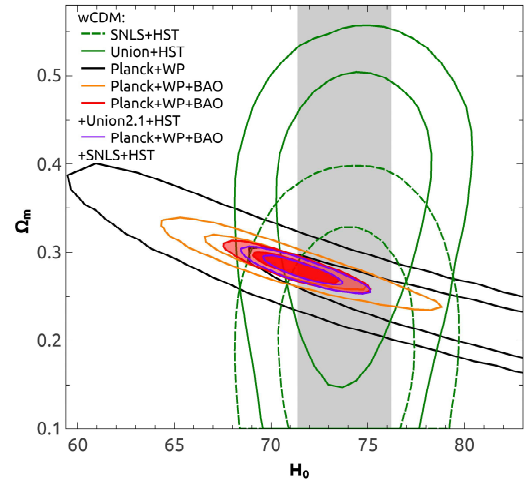}}
\caption{(color online). Contour plots of $\Omega_m - H_0$ in the $w$CDM model with $68\%$ and $95\%$ confidence regions. The black, green, olive solid and olive dashed contours enclose the $68\%$ and $95\%$ confidence regions from Planck+WP, Planck+WP+BAO, Union2.1+HST and SNLS+HST in the $w$CDM model. The red shaded region and yellow contours are the $68\%$ and $95\%$ confidence regions from the combinations Planck+WP+BAO+Union2.1+HST and Planck+WP+BAO+SNLS+HST respectively. }

\label{fig:wcdm}
\end{figure}

We switch to the $w$CDM model in which the dark energy equation of state $w$ is assumed to be a constant in this subsection. Once the free parameter $w$ is taken into account, the confidence regions compared to those in the base $\Lambda$CDM model are significantly enlarged. The constraints on $\Omega_m$ and $H_0$ from Planck+WP, Planck+WP+BAO, Union2.1+HST and SNLS+HST are showed in Fig.~\ref{fig:wcdm}. Here the priors we adopt for the parameters are quite large for enclosing the contours, but only the small patches we are interested in are plotted.   

\begin{figure}[ht]
\centerline{\includegraphics[width=\figurewidth]{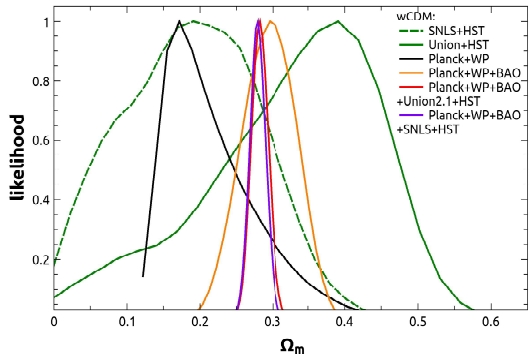}}
\caption{(color online). Distribution of $\Omega_m$ in the $w$CDM model. The black, green, olive solid, olive dashed, red and yellow curves indicate the distributions of $\Omega_m$ from Planck+WP, Planck+WP+BAO, Union2.1+HST, SNLS+HST, Planck+WP+BAO+Union2.1+HST, Planck+WP+BAO+SNLS+HST in the $w$CDM model respectively.}
\label{fig:wcdmm}
\end{figure}

From Fig.~\ref{fig:wcdm}, one can see that the confidence region of Planck+WP (the black contour) is quite large, and it implies that Planck+WP is consistent with both Union2.1+HST and SNLS+HST in the $w$CDM model. Combining with BAO,  the confidence region of CMB data is significantly narrowed. See the green contour in Fig.~\ref{fig:wcdm} which shows that Planck+WP+BAO is still consistent with the SNIa datasets nicely. In this model, one can also find that Planck+WP(+BAO) provides a very tight constraint on $\Omega_m$. But the constraint on $\Omega_m$ from SNIa datasets are broadened, and then there is no inconsistency between Planck+WP(+BAO) and SNIa datasets. See Fig.~\ref{fig:wcdmm} in detail. 

\begin{figure}[ht]
\centerline{\includegraphics[width=\figurewidth]{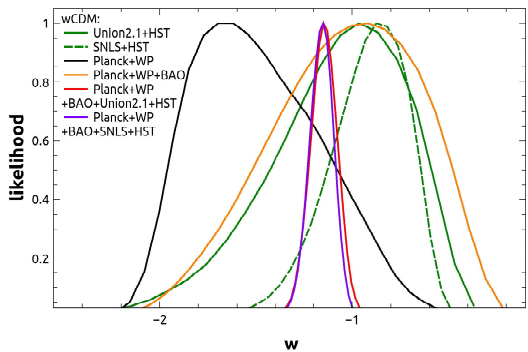}}
\caption{(color online). Distribution of $w$ in the $w$CDM model. The black, green, olive solid, olive dashed, red and yellow curves indicate the distributions of $w$ from Planck+WP, Planck+WP+BAO, Union2.1+HST, SNLS+HST, Planck+WP+BAO+Union2.1+HST, Planck+WP+BAO+SNLS+HST in the $w$CDM model respectively. }
\label{fig:wcdmw}
\end{figure}

\begin{figure}[ht]
\centerline{\includegraphics[width=\figurewidth]{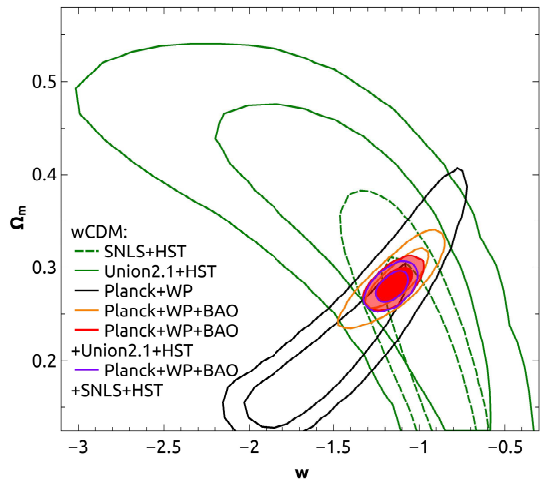}}
\caption{(color online). Contour plot of $\Omega_m - w$ in the $w$CDM model with $68\%$ and $95\%$ confidence regions.  The black, green, olive solid and olive dashed contours enclose the $68\%$ and $95\%$ confidence regions from Planck+WP, Planck+WP+BAO, Union2.1+HST and SNLS+HST in the $w$CDM model. The red shaded region and yellow contours are the $68\%$ and $95\%$ confidence regions from the combinations Planck+WP+BAO+Union2.1+HST and Planck+WP+BAO+SNLS+HST respectively.}
\label{fig:wcdmmw}
\end{figure}

Since the dark energy is not a cosmological constant, we are also interested in the property of dark energy in the $w$CDM model. The distributions of $w$ are showed in Fig.~\ref{fig:wcdmw}. 
Here the data from CMB and BAO is consistent with SNIa datasets, and then we can combine all of them together. 
The constraints on the dark energy equation of state $w$ and $\Omega_m$ are given by 
\m
w=-1.15\pm 0.07, 
\n
\m
\Omega_m=0.283\pm 0.012,
\n
at $68\%$ CL from Planck+WP+BAO+Union2.1+HST; and 
\m
w=-1.16\pm 0.06, 
\n
\m
\Omega_m=0.279_{-0.010}^{+0.011}, 
\n
at $68\%$ CL from Planck+WP+BAO+SNLS+HST. We see that a phantom-like dark energy is preferred at more than $2\sigma$ level. See the contours of $\Omega_m - w$ in Fig.~\ref{fig:wcdmmw} as well. 

We can also add the component of dark radiation. For the combination of Planck+WP+BAO+Union2.1+HST, the constraint on $\Delta N_{\rm eff}$ is $\Delta N_{\rm eff}=0.379_{-0.318}^{+0.314}$
at $68\%$ CL, and $\Delta \chi^2=-1.24$ compared to that without dark radiation. Similarly, for  the combination of Planck+WP+BAO+SNLS+HST, the constraint on $\Delta N_{\rm eff}$ is $\Delta N_{\rm eff}=0.298_{-0.280}^{+0.283}$ at $68\%$ CL, and $\Delta \chi^2=-1.1$ compared to that without dark radiation. We conclude that there is no evidence for dark radiation in the $w$CDM model.

\subsection{(CPL)CDM (+$\Delta N_{\rm eff}$) model}

In the former subsection we focus on the $w$CDM model where $w$ is assumed to be a constant. However the dark energy model with a constant equation of state has less physical interest. 
Here, we focus on the dynamical dark energy model in which the dark energy equation of state is given by 
\m
w(a)=w_0+w_a (1-a), 
\n
where both $w_0$ and $w_a$ are constants, and $w_a$ measures the time evolution of dark energy density. This is a famous parametrization, so-called CPL parametetrization \cite{cpl}, and this model is called by (CPL)CDM model. 

\begin{figure}[ht]
\centerline{\includegraphics[width=\figurewidth]{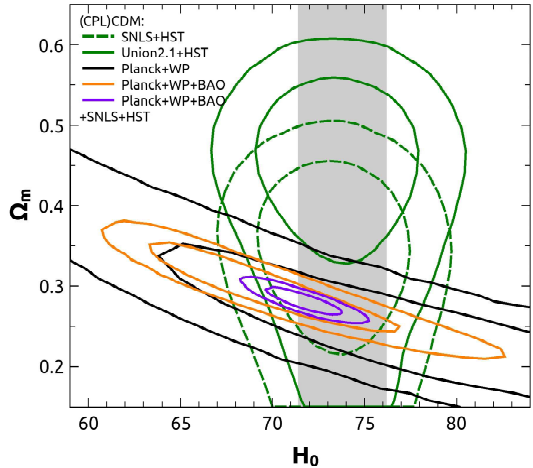}}
\caption{(color online). Contour plot of $\Omega_m - H_0$ in (CPL)CDM model with $68\%$ and $95\%$ confidence regions. The light grey band indicates the $H_0$ prior from Hubble Space Telescope (HST) project at $1\sigma$ level. The black, green, olive solid, olive dashed and yellow contours enclose the $68\%$ and $95\%$ confidence regions from Planck+WP, Planck+WP+BAO, Union2.1+HST and SNLS+HST and Planck+WP+BAO+SNLS+HST in the (CPL)CDM model respectively. }

\label{fig:cplcdm}
\end{figure}

\begin{figure}[ht]
\centerline{\includegraphics[width=\figurewidth]{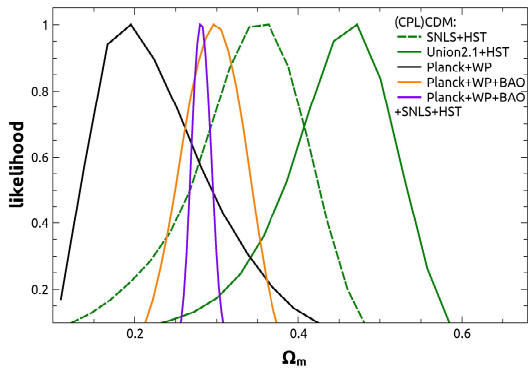}}
\caption{(color online). The distribution of $\Omega_m$ in (CPL)CDM model. The black, green, olive solid, olive dashed and yellow curves indicate the distributions of $\Omega_m$ from Planck+WP, Planck+WP+BAO, Union2.1+HST, SNLS+HST, Planck+WP+BAO+SNLS+HST in (CPL)CDM model respectively.}

\label{fig:cplcdmm}
\end{figure}

\begin{figure}[ht]
\centerline{\includegraphics[width=\figurewidth]{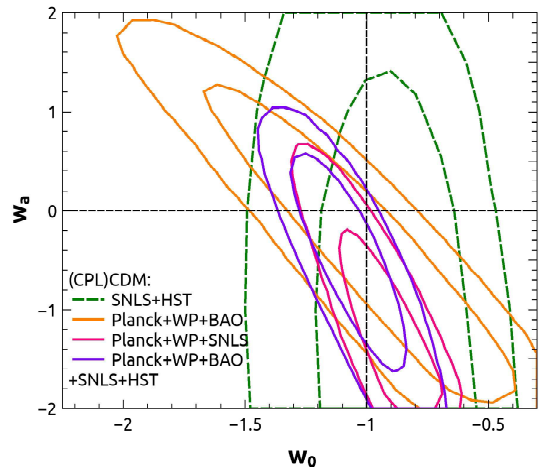}}
\caption{(color online). Contour plot of $w_0 - w_a$ in the (CPL)CDM model. The green, olive dashed, pink and yellow contours enclose the $68\%$ and $95\%$ confidence regions from Planck+WP+BAO, SNLS+HST, Planck+WP+SNLS and Planck+WP+BAO+SNLS+HST in the (CPL)CDM model. }
\label{fig:cplcdmw0wa}
\end{figure}

The constraints on $\Omega_m$ and $H_0$ from Planck+WP, Planck+WP+BAO, Union2.1+HST, SNLS+HST are showed in Fig.~\ref{fig:cplcdm}. 
We see that both Planck+WP and Planck+WP+BAO are consistent with SNLS+HST nicely, but the constraint on $\Omega_m$ from Union2.1+HST is quite large, namely 
\m
\Omega_m=0.425_{-0.076}^{+0.080},
\n
at $68\%$ CL which is discrepant with Planck+WP estimate $(\Omega_m=0.250_{-0.112}^{+0.080};\ \hbox{at 68}\%\ \hbox{CL})$ at about $2.3\sigma$. See Fig.~\ref{fig:cplcdmm} in detail. 
Considering that Planck+WP+BAO is consistent with SNLS+HST in (CPL)CDM model, we combine all of them to do the data fitting, and the results are showed in Figs.~\ref{fig:cplcdm}, \ref{fig:cplcdmm} and \ref{fig:cplcdmw0wa}.
In particular, from Fig.~\ref{fig:cplcdmw0wa}, a cosmological constant is disfavoured at around $2\sigma$ for Planck+WP+SNLS; but still stays within $2\sigma$ confidence region for Planck+WP+BAO+SNLS+HST. Since $w_a=0$ is included in the $1\sigma$ confidence region, there is no evidence for a dynamical dark energy.

Similar to the previous section, we can add the component of dark radiation in the (CPL)CDM model. 
For  the combination of Planck+WP+BAO+SNLS+HST, the constraint on $\Delta N_{\rm eff}$ is $\Delta N_{\rm eff}=0.218_{-0.307}^{+0.314} $ at $68\%$ CL, and $\Delta \chi^2=-0.84$ compared to that without dark radiation. We conclude that there is no evidence for dark radiation as well.

\section{Discussion}

Even though the CMB datasets (including Planck and WP) are consistent with BAO in the base $\Lambda$CDM model, 
there are some tensions between the CMB datasets and the SNIa datasets, such as HST and SNLS where a larger $H_0$ and less matter are preferred. Adding the dark radiation can relax the tension on $H_0$ between Planck and HST, $\Omega_m$ obtained by Planck is still not nicely compatible with that from SNLS. In the $\Lambda$CDM+$\Delta N_{\rm eff}$ model, there is an overlap region in the plot of $\Omega_m - H_0$ for Planck+WP, BAO, Union2.1 and HST. Combining all of them, we find that the dark radiation is preferred at $2.4\sigma$ level.

In the $w$CDM model, due to the broadened confidence region for $\Omega_m$ and $H_0$, Planck+WP, BAO, Union2.1, SNLS and $H_0$ from HST are consistent with each other, and then there is not any tension any more. However, the price we need to pay is that a phantom-like dark energy is called for. In this model we did not find any evidence for existence of a dark radiation. 

Once a time-evolving dark energy model ((CPL)CDM model) is taken into account, the data of Union2.1 prefers a quite large $\Omega_m$ which has a tension with CMB data (Planck+WP) at around $2.3\sigma$ level. Combining Planck+WP, BAO, SNLS and $H_0$ from HST, we find that a cosmological constant is disfavoured at more than $1\sigma$ level, but still stays inside the $2\sigma$ level. 

Comparing $\Lambda$CDM+$\Delta N_{\rm eff}$ to $w$CDM model, both of them have the same number of parameters, and the combination of of Planck+WP+BAO+Union2.1+HST does not prefer any of them ($|\Delta \chi ^2|=0.2$), but $w$CDM model provides a much better fitting for the combination of Planck+WP+BAO+SNLS+HST. 
Compared to $w$CDM model, the time-evolving dark energy model ((CPL)CDM) has one more parameter, but it only improves the fitting result by $\Delta\chi^2=-0.3$ which indicates that there is no statistical evidence for a time-evolving dark energy model. To summarize, only the $w$CDM model does significantly relax the tensions between Planck and other astrophysical datasets and the preferred $H_0$ and $\Omega_m$ are showed in Fig.~\ref{fig:wcdm} and Table.~\ref{tab:sum}.


See some other approaches in \cite{Zhang:2013hma,Li:2013dha,Marra:2013rba}. For more discussions about the dynamical dark energy and dark radiation see, for example, \cite{Zhao:2012aw,Nojiri:2013ru,DiBari:2013dna,Li:2012via,Zhang:2012jsa}.

\vspace{5mm}
\noindent {\bf Acknowledgments}

We would like to thank X.~D.~Li for useful discussions. 
We acknowledge the use of Planck Legacy Archive and ITP for providing computing resources.
This work is supported by the project of Knowledge Innovation Program of Chinese Academy of Science and grants from NSFC (grant NO. 10821504, 11322545 and 11335012).



\end{document}